\documentstyle[12pt,frascatiphys]{article}
\begin{document}
\title{ 
DOES THE LIGHT AND BROAD $\sigma(500)$ EXIST?
}
\author{Nils A T\"ornqvist        \\
{\em Physics Dept. University of Helsinki, PB9, Fin-00014 Helsinki, Finland} \\
}
\maketitle
\baselineskip=14.5pt
\begin{abstract}
The  lightest scalar and pseudoscalar nonets are discussed within the framework
of the broken U3$\times$U3 linear sigma model, and it is shown that already 
at the tree level this model
 works remarkably well predicting scalar masses and 
couplings not far  from present experimental values, 
when all parameters are fixed from the 
pseudoscalar masses and decay constants. The linear $\sigma$ model
is the simplest way to
implement chiral symmetry together with the broken 
SU3 of the quark model, and its success in understanding 
experiment is comparable to that of
the naive quark model for the heavier multiplets. It is argued that this
strongly suggests that the light and very broad $\sigma$ 
resonance exists near 500 MeV.\end{abstract}
\baselineskip=17pt

\section{Introduction}

\def \gam {\frac{ N_f N_cg^2_{\pi q\bar q}}{8\pi} }
\def \gamm {N_f N_cg^2_{\pi q\bar q}/(8\pi) }
\def \be {\begin{equation}}
\def \ba {\begin{eqnarray}}
\def \ee {\end{equation}}
\def \ea {\end{eqnarray}}
\def \gap {{\rm gap}}
\def \gaps {{\rm {gaps}}}
\def \gappp {{\rm \overline{\overline{gap}}}}
\def \im {{\rm Im}}
\def \re {{\rm Re}}
\def \Tr {{\rm Tr}}
\def \P {$0^{-+}$}
\def \S {$0^{++}$}
\def \uu {$u\bar u+d\bar d$}
\def \ss {$s\bar s$}

In this talk\footnote{Invited talk at 
the Workshop on Hadron Spectroscopy, March 8-13, 1999 Frascati, Italy. 
To appear in the Frascati Physics Series.} 
I shall discuss mainly the light and 
broad $\sigma$, which was picked by Matts Roos and myself\cite{NAT}
from the particle data group wastebasket 4 years ago, having been there for 
over 20 years. Today an increasing number of papers, 
many of which have been reported at this meeting\cite{sighere},
  are quoting its 
parameters, with a pole position near 
500-i250 MeV (See the table 1).

An important question today is: Does this broad resonance really exist? 
And if so, what is its nature, together with the other light scalar mesons in 
the 1 GeV region? The naive quark model (NQM), which works reasonably 
well for the vectors and heavier multiplets, 
definitely fails for the scalars taken as
$\sigma(500), f_0(980), a_0(980)$ and $K^*_0(1420)$.

\begin{table}
\centering
\caption{ \it $\sigma$ pole position.
}
\vskip 0.1 in
\begin{tabular}{|l|c|} \hline
Reference          &  pole position (MeV \\
\hline
\hline
Kaminski et al.\cite{sighere}& $532\pm 12-i(259\pm 7)$  \\
 Locher et al.\cite{sighere}  & $424-i213$ \\ 
 Harada  et al.\cite{sighere} & $\approx 500-i250$   \\
 Lucio et al.\cite{sighere}   & $\approx 400-i200$   \\
 Ishida et al.\cite{pdg98}   & $ 602\pm26-i(196\pm27)$    \\
 Kaminski et al.\cite{pdg98}  & $537\pm20-i(250\pm 17)$   \\ 
 Oller    et al.\cite{pdg98}  & $469.5-i178.6$  \\
 T\"ornqvist   et al.\cite{pdg98}  & $470-i250$   \\
 Amsler    et al.\cite{pdg98}  & $1100-i300$   \\
 Amsler    et al.\cite{pdg98}  & $400-i500$   \\
 Janssen   et al.\cite{pdg98}  & $387-i305$   \\
 Achasov   et al.\cite{pdg98}  & $525-i269$   \\
 Zou       et al.\cite{pdg98}  & $370-i356$   \\
 Zou       et al.\cite{pdg98}  & $408-i342$   \\
 Au        et al.\cite{pdg98}  & $870-i370$   \\
 Beveren   et al.\cite{pdg98}  & $470-i208$   \\
 Estabrooks\cite{pdg98}        & $750\pm50-i(450\pm 50)$   \\
 Protopescu et al.\cite{pdg98} & $660\pm100-i(320\pm 70)$   \\
 Basdevant et al.\cite{pdg98}  & $650-i370$  \\
 Scadron  et al.\cite{scad}& $\approx 500-i250$   \\ 
 Lucio et al.\cite{lucio}    &$600^{+200}_{-100}-i350 $\\
 Igi et al.\cite{igi}&$\approx 760$\\
\hline
\end{tabular}
\label{extab}
\end{table}
 
We all believe the vectors ($\rho, \omega,\phi,K^*)$ and heavier well 
established multiplets
are $q\bar q$ states because with a few parameters,
 such as an equally spaced bare mass spectrum, a small OZI rule violating 
parameter and SU3$_f$ related coupligs, we can describe
the masses, widths and couplings of the whole nonet. 
If we had data only on the $\rho(770)$ 
we could not conclude that it is $q\bar q$. But with the 
successful SU3$_f$ predictions for the whole nonet we strongly believe it is 
$q\bar q$.

The same is true for the $\sigma$. No single analysis of the $\pi\pi$
S-wave, however refined, could ever decide on what is the nature of the 
$\sigma$.
Even  the decision as to 
whether it really exists, cannot be done using data on the 
$\pi\pi$ S-wave alone, since there are inherent, model dependent, 
ambiguities as to how to continue analytically to a pole which is
far from the physical region, as is the case for the broad $\sigma$.

It is also obvious why the NQM fails for the scalars:
Chiral symmetry is absent in the NQM, but is crucial for the scalars.
Chiral symmery is believed to be broken in the vacuum, and
 two of the scalars ($\sigma$ and $f_0$) have the same quantum 
numbers as the vacuum. 
Thus to understand the scalar nonet in the same way as we 
believe we understand the vectors, 
and to make a sensible comparison with experiment, 
one must include chiral symmetry in 
addition to flavour symmetry in the quark model. 
The simplest such chiral quark model is the 
linear U3$\times$U3 sigma model with 3 flavours.    

Then we can treat both the scalar and pseudoscalar nonets simultaneously, and
on the same footing, getting automatically small masses for the pseudoscalar 
octet,
and symmetry breaking through the vacuum expectation values (VEV's) of the 
scalar fields.

As an extra bonus we have in principle a 
renormalizable theory, i.e. ``unitarity 
corrections'' are calculable.  
In fact, in the flavour symmetric limit the unitarity corrections can 
be thought to be already included 
into the mass parameters of the theory, once the original 4-5 parameters are
replaced by the 4 physical 
masses for the singlet and octet $0^{-+}$ and $ 0^{++}$ masses 
and the  $\sigma$ VEV. 

Unfortunately this over 30 years old 
model\cite{sigma} has had very few phenomenological 
applications. An important exception is the intensive efforts of 
M. Scadron and collaborators. %

\section{The Linear sigma model with 3 flavours}

The well known linear sigma model\cite{sigma} generalized to 3 flavours
 with complete  scalar ($s_a$) and pseudoscalar ($p_a$)
nonets has at the tree-level the 
Lagrangian  the same flavour and chiral symmetries 
as massless QCD. The  U3$\times$U3 Lagrangian with a symmetry breaking term 
${\cal L}_{SB}$ is  
\be 
 {\cal L}=
\frac 1 2 \Tr [\partial_\mu\Sigma \partial_\mu\Sigma^\dagger]
-\frac 1 2 \mu^2\Tr [\Sigma \Sigma^\dagger] -\lambda \Tr[\Sigma\Sigma^\dagger
\Sigma\Sigma^\dagger]\ -\lambda' 
(\Tr[\Sigma\Sigma^\dagger])^2+{\cal L}_{SB} \ .
\label{lag}
\ee

Here   $\Sigma$ is a $3\times 3$ complex
matrix, $\Sigma=S+iP= \sum_{a=0}^8(s_a+ip_a)\lambda_a/\sqrt 2$, in which
$\lambda_a$ are the Gell-Mann matrices, normalized as $\Tr[\lambda_a\lambda_b]=
2\delta_{ab}$, and where for the singlet 
$\lambda_0 = (2/N_f)^{1/2} {\bf 1}$ is 
included. Each  meson in Eq. (1) has a definite SU3$_f$ symmetry
content, which in the quark model means that it has the same  flavour
structure as
a  $q\bar q$ meson. Thus the fields $s_a$ and $p_a$ and 
potential terms in Eq.~(1) can be given
 a conventional quark line structure\cite{NATPL}. 


The symmetry breaking terms are most simply:
\be
{\cal L}_{SB}=\epsilon_\sigma \sigma_{u\bar u+d\bar d} + 
\epsilon_{s\bar s} \sigma_{s\bar s} +c[\det \Sigma +\det \Sigma^\dagger]\ ,
\ee 
which  give the pseudoscalars mass and  break the 
flavour and $U_A(1)$ symmetries. The small parameters $\epsilon_i$ 
can be expressed in terms of 
the pion and kaon decay constants and masses:
 $\epsilon_\sigma = m_\pi^2f_\pi$, $\epsilon_{s\bar s}=
(2m_K^2f_K-m_\pi^2f_\pi)/\sqrt 2$.

My fit to the scalars with the unitarized quark model (UQM)\cite{NAT} 
is essentially a  unitarization of 
eq.\ref{lag} with $\lambda\approx 16$ and $\lambda '= 0$, 
and with the main symmetry breaking generated by 
putting the pseudoscalar masses  at their physical values.
The  model was used as an effective theory 
with a symmetric smooth 3-momentum cutoff 0.54 GeV/c given by a gaussian 
form factor. Such a form factor is natural, since physical mesons are of 
course not pointlike, but have finite size of 0.7-0.8 fm.
The fit  included the Adler zeroes which follow from eq.1,
 but only approximate crossing symmetry.  

Eq.~(1) without ${\cal L}_{SB}$ 
is clearly invariant under $\Sigma \to U_L\Sigma U_R^\dagger$ of
U3$\times$U3.
After  shifting the flavourless scalar fields by  
the VEV's ($\Sigma \to \Sigma+ V$) to the minimum of the potential, 
the scalars aquire masses and also 
the pseudoscalars obtain a  (small) mass because of  
${\cal L}_{SB}$. Then the $\lambda$ and $\lambda'$ terms  
generate trilinear $spp$ and $sss$
couplings, in addition to those coming from the $U_A(1)$ symmetry breaking 
determinant term. The $\lambda$ term, which turns out to be the largest, 
obeys the OZI rule, while the small $\lambda '$ term, and $c$ violate this rule. 
 
\section{ Tree-level masses and couplings.}

It is an ideal problem for a symbolic program like Maple V to calculate the
predicted masses, and couplings from the Lagrangian, which has 
6 parameters $\mu,\lambda,\lambda',c$ 
and the VEV parameters $u=d$ and $s$, which define a diagonal 
matrix for the flavourless meson VEV's: $V=diag[u,d,s]$. These are at the 
tree level related to the pion and kaon decay constants through
 $u=d=<\sigma_{u\bar u+d\bar d}>/\sqrt 2=f_\pi/\sqrt 2$ (assuming 
isospin exact) and $s=<\sigma_{s\bar s}>=(2f_K-f_\pi)/\sqrt 2$: 
One finds  denoting the often occurring combination 
$\mu^2+4\lambda'(u^2+d^2+s^2)$ by $\bar \mu^2$, and expressing the flavourless mass matrices in the ideally mixed frame:
\eject
\ba
m^2_{\pi^+}\ &=&\bar \mu^2 + 4\lambda(u^2+d^2-ud)+2cs \cr
m^2_{K^+}  \ &=&\bar \mu^2 + 4\lambda(u^2+s^2-su)+2cd \cr
\begin{array}{c} m^2_\eta \\ m^2_{\eta'} \end{array} &=& diag
\left( \begin{array}{cc} \bar \mu^2+2\lambda(u^2+d^2)-2cs & -c\sqrt 2 (u+d) \\
-c\sqrt 2 (u+d)& \bar \mu^2+4\lambda s^2 \end{array}\right) 
\\
m^2_{a_0^+}\   &=&\bar \mu^2 + 4\lambda(u^2+d^2+ud)-2cs \cr
m^2_{\kappa^+} \  &=&\bar \mu^2 + 4\lambda(u^2+s^2+su)-2cd \cr
\begin{array}{c} m^2_\sigma \\ m^2_{f_0} \end{array} &=& diag
\left( \begin{array}{cc} \bar \mu^2+4\lambda'(u+d)^2+6\lambda(u^2+d^2)+2cs
 & (4\lambda' s+c)\sqrt 2 (u+d) \\
(4\lambda' s+c)\sqrt 2 (u+d)& \bar \mu^2+
8\lambda' s^2+12\lambda s^2 \end{array}\right) \nonumber
\cr
\ea
\begin{table}
\centering
\caption{ \it Predicted masses in MeV and mixing angles for two values of the 
 $\lambda'$ parameter. The 
 asterix means that $m_\pi,m_K$
and $m_\eta^2+m^2_{\eta'}$ are fixed by experiment together with $f_\pi$ and 
$f_K$.  }
\vskip 0.1 in
\begin{tabular}{|l|c|c|c|} \hline
Quantity       &  Model $\lambda '=1$& Model $\lambda '=3.75$& Experiment \\
\hline
\hline
$m_\pi$   &  137$^{*)}   $& 137$^{*)}   $ &137  \\
$m_K$     &  495$^{*)}   $& 495$^{*)}   $ &495  \\
$m_\eta  $&  538$^{*)}   $& 538$^{*)}   $ &547.3  \\
$m_{\eta'}$ &963$^{*)}   $& 963$^{*)}   $ &957.8  \\ 
$\Theta^{\eta'-singlet}_P$ &-5.0$^\circ$ &-5.0$^\circ$ &(-16.5$\pm6.5)^\circ$\cite{pdg98} \\ 
$m_{a_0}$ &1028 &1028 &  983  \\ 
$m_{\kappa}$ &1123 &1123&   1430  \\ 
$m_{\sigma}$ &651 &619  & 400-1200  \\ 
$m_{f_0}$ &1229 &1188   &980  \\ 
$\Theta^{\sigma-singlet}_S$ 
& 21.9$^\circ$ & 32.3$^\circ$ &(28-i8.5)$^\circ$\cite{NAT}  \\ 
\hline
\end{tabular}
\label{extab2}
\end{table}

 Now we fix 5 of the 6 parameters except $\lambda'$ by the 5 experimental 
quantities
$m_\pi$, $m_K$, $m^2_\eta+m^2_{\eta'}$,
$f_\pi=92.42$ MeV and $f_K=113$ MeV\cite{pdg98}. One finds at the tree level 
$\lambda=11.57$, 
$\bar \mu^2=0.1424$ GeV$^2$, $c=-1701$ MeV, $u=d=65.35 $MeV 
and $s=94.45$MeV.
The remaining $\lambda'$ paramerer affects only the $\sigma$ and $f_0$ masses 
and their trilinear 
couplings. This dependence is rather weak for the masses, but is very sensitive
to the  couplings as it changes the ideal mixing angle for the scalars. 
It turns out below that this must be small to fit the tri-linear couplings. 
By putting $\lambda'=1$ one gets a reasonable compromise for most of 
these couplings. With $\lambda'\approx 3.75$ one almost 
cancels the OZI rule breaking 
coming from the determinant  term, and the scalar mixing becomes near ideal
(for $\lambda '=-c/(4s)=4.5$ the cancellation is  exact).

As can be seen from Table 2 the predictions 
 are not far from the experimental values. 
Considering that one expects from our previous analysis\cite{NAT}
 that unitarity 
corrections can easily be more than 20\% , and should 
go in the right direction, one must
conclude that these results are even better than expected.

The trilinear coupling constants follow from the Lagrangian,
and are at the tree level:
\ba
g_{\sigma\pi^+\pi^-}&=&\cos \phi_S^{id}(m^2_\sigma-m^2_\pi)/f_\pi \cr
g_{\sigma K^+K^-}&=&-\sqrt 3\sin( \phi_S^{id}-35.26^\circ )
(m^2_{\sigma}-m^2_K)/(2f_K) \cr
g_{f_0\pi^+\pi^-}&=&\sin \phi_S^{id}(m^2_{f_0}-m^2_\pi)/f_\pi \cr
g_{f_0 K^+K^-}&=&\sqrt 3\cos( \phi_S^{id}-35.26^\circ )
(m^2_{f_0}-m^2_K)/(2f_K) \\
g_{a_0\pi\eta}&=&\cos \phi_P^{id}(m^2_{a_0}-m^2_\eta)/f_\pi \cr
g_{a_0\pi\eta'}&=&\sin \phi_P^{id}(m^2_{a_0}-m^2_{\eta'})/f_\pi \cr  
g_{a_0 K^+K^-}&=&(m^2_{a_0}-m^2_K)/f_K \cr
g_{\kappa^+K^0\pi^+}&=&(m^2_\kappa-m^2_\pi)/(\sqrt 2 f_K) \cr
g_{\kappa K^+\eta}&=&-\sqrt 3\sin( \phi_P^{id}-35.26^\circ )
(m^2_{\kappa}-m^2_\eta)/(2f_K) \cr  
g_{\kappa K^+\eta'}&=&\sqrt 3\cos( \phi_P^{id}-35.26^\circ )
(m^2_{\kappa}-m^2_{\eta'})/(2f_K) \ .\nonumber \cr  
\ea
Here $\phi^{id}_P=54.73^\circ+\Theta^{\eta'-singlet}_P$ i.e.
the angle between $s\bar s$ and $\eta'$
and $\phi^{id}_S=-35.26^\circ+\Theta^{\sigma-singlet}_S 
$, 
i.e. the angle between $u\bar u+d\bar d$ and  $\sigma$.

\begin{table}
\centering
\caption{ \it Predicted couplings $\sum_i\frac{g_i^2}{4\pi}$ (in GeV$^2$)
, when $\lambda'=1$, 
compared with experiment and predicted widths with experiment (in MeV). (We 
have used isospin invariance to get the sum over charge channels, when data is
for one channel only.)   The 
predicted $f_0\to \pi\pi$ width is extremely
 sensitive to the value of $\lambda^\prime$ (for $\lambda '= 
3.75 $  it nearly vanishes)
and unitarity effects as discussed in the text. }
\vskip 0.1 in
\begin{tabular}{|l|c|c|c|c|} \hline
Process       &  $ \sum_i\frac{g_i^2}{4\pi}$ &  $ \sum_i\frac{g_i^2}{4\pi}$  &
$\sum_i\Gamma_i$ & $\sum_i\Gamma_i$ \\
      & in model &in experiment & model&experiment\\ 
\hline
\hline
$\kappa^+\to K^0\pi^++K^+\pi^0$   &  7.22 & -  & 678 & $278\pm23$\cite{pdg98}   \\
$\kappa^+\to K^+\eta $     &  0.28 & $\approx$ 0 & & \\
$\sigma\to \pi^+\pi^-+\pi^0\pi^0$   &  2.17 & 1.95\cite{ach} & 574 
&300-1000\cite{pdg98} \\
$\sigma\to K^+K^-+K^0\bar K^0$   &  0.16 & 0.004\cite{ach}   & 0   & 0\\
$f_0\to    \pi^+\pi^-+\pi^0\pi^0$   
&  1.67 & 0.765$^{+0.20}_{-0.14}$\cite{nov} & see text & 40 - 100\cite{pdg98}\\
$f_0\to    K^+K^-+K^0\bar K^0$   &  6.54 & 4.26$^{1.78}_{-1.12}$\cite{nov}   & 0   & 0\\
$a_0^+\to \pi^+\eta$   &  2.29 & 0.57\cite{nov}  & 273 see text & 50 - 100\cite{pdg98}\\
$a_0^+\to K^+\bar K^0$ &  2.05 &  1.34$^{+0.36}_{-0.28}$\cite{nov} &  0 &0  \\
\hline
\end{tabular}
\label{extab2}
\end{table}

In some of the channels of  table 3 the resonance is 
below threshold and the widths vanish at the resonance mass. 
However, the coupling
constants have then been determined through a loop diagram from
$\phi\to\gamma\pi\pi$ and $\phi\to\gamma\pi\eta$ (albeit in a somewhat model 
dependent way) by the Novosibirsk group.
For channels where the phase space is large, it is important that one includes
a form factor related to the finite size of physical mesons. In the quark 
pair creation ($^3P_0$) model a radius of 0.8 fm 
leads to a gaussian form factor,
as in the formula below, where $k_0 \approx 0.56$ GeV/c (as was found
 in the UQM\cite{NAT}). 
Thus the  widths are computed from the formula: 
\be
\Gamma(m)=\sum_{isospin}\frac{g_i^2}{8\pi}
\frac{k^{cm}(m)}{m^2}e^{-[k^{cm}(m)/k_0]^2}\ .
\ee

As can be seen from table 3 most of the  couplings are not far from experiment.
The main exception is the $f_0\to\pi\pi$ coupling and width, but this is 
extremely sensitive to the ideal mixing angle. If one choses 
$\lambda'= 3.75$  this mixing angle nearly vanishes 
($\phi^{id}_S=-3.0^\circ$) together with the $f_0\to\pi\pi$ coupling 
(c.f. eq.4).
From our experience with the UQM\cite{NAT} the couplings, when unitarized,
 are very sensitive to especially 
the nearby $K\bar K$  threshold. Similarily the $a_0\to\pi\eta$ peak width
is reduced, because of
 the $K\bar K$ theshold, by up to a factor 5. Therefore one cannot 
expect that the tree level couplings should agree better with data 
than what those of Table 3 do. In fact, I was myself astonished by the fact
that the agreement turned out to be this good. After all,
 this is a very strong
coupling model ($\lambda=11.57$, leading to large $g_i^2/4\pi$)
and higher order effects should be important.

\section{Conclusions.}
In summary, I find that the linear sigma model 
with three flavours works, at the tree level, much  
better than expected. It
 works, in my opinion, just as well as the 
naive quark model works for the heavier nonets.
 One should of course include higher order effects, i.e., the model
 should be unitarized phenomenologically, 
e.g., along the lines of the UQM\cite{NAT}, 
whereby a more detailed data comparison becomes meaningful.

Those working on chiral perturbation theory and nonlinear sigma models
 usually point out that
 the linear model does not predict all low energy constants 
correctly. However, one should remember that the energy regions of validity
are different for the two approaches.
 Chiral perturbation theory usually breaks down when one approaches
the first scalar resonance. 
The  linear sigma model, on the other hand,
includes the scalars from the start 
and can be a reasonable interpolating model in the intermediate 
energy region near 1 GeV, where QCD is too difficult to solve.

These results strongly favour the interpretation that the $\sigma(500)$, $
a_0(980)$, $f_0(980)$, $ K^*_0(1430)$  belong to the same nonet, 
and that they are the chiral partners of the $\pi$, $\eta$, $ K$, $\eta '$. 
If the latter are believed to be unitarized $q\bar q$ states, 
so are the light scalars $\sigma(500),$ $
a_0(980),$ $ f_0(980),$ $ K^*_0(1430)$, and the broad 
$\sigma(500)$ should be interpreted as an existing resonance.

The $\sigma$ is a very important hadron indeed, 
as is evident in the sigma model, because this is the boson
which gives the constituent quarks most of their mass
and thereby it gives also 
the light hadrons most of  their mass. It is the
Higgs boson of strong interactions. 

\end{document}